\documentclass[12pt]{article}
   \newcommand{\nl}{\nonumber \\}
   \newcommand{\beq}{\begin{equation}}
   \newcommand{\eeq}{\end{equation}}
   \newcommand{\beqa}{\begin{eqnarray}}
   \newcommand{\eeqa}{\end{eqnarray}}
   
   \newcommand{\bnab}{\mbox{\boldmath ${\nabla}$}}

   \newcommand{\dadb}[2]{\frac{{  d}#1}{{  d}#2}}

   \newcommand{\bu}{{\mathbf u}}
   \newcommand{\bv}{{\mathbf v}}

   \newcommand{\bA}{{\mathbf A}}
   \newcommand{\bB}{{\mathbf B}}

   \newcommand{\bE}{{\mathbf E}}

   \newcommand{\bJ}{{\mathbf J}}

\newcommand{\ed}{\end{document}}
\usepackage{color}
\usepackage{epsfig}
\usepackage{epstopdf}
\usepackage[a4paper,
            bindingoffset=0.2in,
            left=1in,
            right=1in,
            top=1in,
            bottom=1in,
            footskip=.25in]{geometry}
\usepackage{hyperref}
\hypersetup{
    colorlinks=true,
    linkcolor=red,
    filecolor=magenta,      
    urlcolor=cyan,
    citecolor = cyan,
    }
\begin{document}

\begin{center}
\noindent
{\large \bf Hybrid fluid-kinetic cylindrical equilibria  \\ \vspace{2mm} 
with axial background magnetic field}\vspace{4mm}

D. A. Kaltsas$^{1}$, A. I. Kuiroukidis$^1$,  and G. N. Throumoulopoulos$^1$ \vspace{4mm}

$^1$Department of Physics, University of Ioannina, GR 451 10 Ioannina, Greece  \vspace{4mm}

\vspace{3mm}
Emails: kaltsas.d.a@gmail.com, \  a.kuirouk@uoi.gr,\   gthroum@uoi.gr
\end{center}

%%%%%%%%%%%%%%%%%%%%%%%%%%%%%%%%%%%%%%%%%%%%%%%%%%%%%%%%%%%%%%%%%%%%%%%%%%%%%%%%%%%%%%%%%%%%%%%%%%%%

\begin{abstract}

 Self-consistent, one-dimensional quasineutral screw-pinch equilibria are constructed within a hybrid model that couples fluid electrons with kinetic ions governed by the Vlasov equation. The equilibria depend on the radial coordinate perpendicular to the cylindrical axis and include an axial background magnetic field. Adopting a three-parameter ion distribution function depending on the
energy and the canonical momenta conjugate to the two ignorable coordinates, the problem
is reduced to a set of four quasilinear ODEs which are solved numerically. Both static equilibria and equilibria with macroscopic ion sheared velocities are obtained. The pressure of
the electron fluid is isotropic and the electron contribution to the current density is parallel
to the magnetic field, while the kinetic ions are associated with a non-gyrotropic pressure
tensor. By means of the solutions the various equilibrium quantities are calculated and the
impact of the free parameters on the equilibrium characteristics is examined.  %{\color{red} These equilibria are relevant to magnetic flux ropes in Earth's %magnetosphere having length scales comparable to the ion inertial length.}
\vspace{1mm}

\noindent Keywords: Hybrid fluid-kinetic models, screw-pinch equilibria, magnetic confinement

\end{abstract}

%\newpage

%%%%%%%%%%%%%%%%%%%%%%%%%%%%%%%%%%%%%%%%%%%%%%%%%%%%%%%%%%%%%%%%%%%%%%%%%%%%%%%%%%%%%%%%%%%%%%%%%%%%

\section{Introduction}\

Energetic particles, i.e. particles that have been accelerated to much higher energies than the average thermal energy, are ubiquitous  in laboratory, space
and astrophysical plasmas \cite{McTu}. Examples are energetic electrons in planets magnetospheres and ionospheres, solar flares, solar corona, cosmic rays and during ELMs in tokamaks as well as energetic ions in coronal mass injections and solar wind \cite{SiJo}. Also, energetic ions are created by  laser plasma interactions \cite{BrJa},  by fusion reactions and by neutral beam injection and radiofrequency heating in magnetic confinement systems \cite{Me,HeSa}.  In particular, we mention the production of energetic $\alpha$-particles in burning fusion plasmas in connection with the ITER project \cite{SaSp}. In addition to fusion reactions and external methods of plasma heating, other physical processes  producing energetic particles include magnetic reconnection, shock waves,  and turbulence.   In particular, magnetic reconnection taking place at current sheets in Earth's magnetotail not only produces energetic particles but also generates magnetic structures known as magnetic flux ropes (MFRs) (e.g. see \cite{Sharma2008,Smith2024}) which are structures similar to the cylindrical screw pinch. In Earth's magnetotail these structures appear as chains of magnetic islands as the magnetic reconnection occurs on multiple X-points on the current sheet as deduced by Cluster observations \cite{Escoubet2015}. 

The energetic particles can affect significantly the plasma equilibrium, drive waves and instabilities or interact with existing instabilities, accelerate other particles  and modify the turbulence and the energy transport in the plasma. Thus, a self-consistent description of plasmas with energetic particles is crucial for modeling these phenomena. Plasmas containing populations with energetic ions can apparently be described in the framework of kinetic theory. However, whenever time and space scales comparable to the ion gyroperiod and ion inertial length are involved, instead of computational expensive complete kinetic treatments, hybrid models are preferable, consisting of a kinetic description of the ions and a fluid description for the electrons. Specifically, in the collisionles regime such models employed the Vlasov equation for the ions, fluid equations for the electrons coupled each other self-consistently with the Maxwell equations  to study pertinent dynamical processes including magnetic reconnection and turbulence \cite{TaMi,VaTr,SeVa,CeCa,FrSe}.  Also, the Hamiltonian structures of several hybrid  models through pressure- and current-coupling schemes were identified in \cite{Tr}. In connection with the present study, one dimensional, plane, hybrid  equilibria with sheared flow, directed either parallel or perpendicular to a uniform magnetic field, were constructed in \cite{MaPe}. In addition, one-dimensional, quasineutral, hybrid equilibrium states in plane geometry were obtained in \cite{KaMo} and axisymmetric hybrid equilibria with applications to tokamak plasmas was the subject of \cite{KaKu}. Furthermore, one-dimensional cylindrical hybrid steady states were constructed in \cite{Ng}  in relation to Bernstein–Greene–Kruskal modes in the presence of a magnetic field, as a possible theoretical explanation of flux-rope structures observed in many regions of space plasmas. In general, hybrid equilibrium states are required as initial conditions of dynamical processes, e.g. waves and instabilities, involving ion space and time scales. Also, from the equilibrium point of view it is interesting to examine how the kinetic ion component affect the equilibrium characteristics.

Aim of the present study is to construct one-dimensional,  screw pinch equilibria by extending those obtained in \cite{KaMo} to cylindrical geometry including a constant, axial magnetic field. As in \cite{KaMo},  the quasineutrality condition is employed in order to express the electrostatic potential in terms of the components of the vector potential.  For the electronic fluid we adopt the Boltzmann equation. Also, since Maxwellian distribution functions do not produce electric currents and shifted Maxwellians can lead to spurious effects \cite{MaPe}, we employ distributions depending on all three constants of motion (i.e. the energy, and the two canonical momenta conjugate to the ignorable cylindrical coordinates) in the particular form introduced in \cite{Ng} and employed in  \cite{KaMo}. 

In Section 2 the hybrid equilibrium model is presented  describing the ions  by the Vlasov equation and the electrons as a massless fluid with constant temperature and isotropic pressure. Then, it is shown that the current is composed by a kinetic ion component and a Beltrami electron component, that is,  having current  parallel to the magnetic field. Also,  the ion pressure tensor is non-gyrotropic.  Subsequently, by adopting cylindrical symmetry,  i.e. assuming that any equilibrium quantity solely depends on the radial cylindrical coordinate, the problem is reduced to a set of four first-order quasilinear ODEs for the axial and azimuthal components  of the magnetic field and the respective components of the vector potential, which are solved numerically. In Section 3 several solutions are presented describing diamagnetic and paramagnetic equilibria as well as equilibria having mixed magnetic properties. Also, the impact of the free parameters,  associated  with the background axial magnetic field, the Beltrami electron fluid,  the radius of the cylindrical cross-section and the non-Maxwellian part of the ion distribution function,  is examined. A summary of the study is given in Section 4.  

%%%%%%%%%%%%%%%%%%%%%%%%%%%%%%%%%%%%%%%%%%%%%%%%%%%%%%%%%%%%%%%%%%%%%%%%%%%%%%%%%%%%%%%%%%%%%%%%%%%%%%%%%%

\section{Hybrid equilibrium equations }\

The hybrid equilibrium model for a plasma with kinetic ions and massless electrons consists of the following set of equations \cite{KaMo,KaKu}:
\begin{eqnarray}
&&\bv\cdot \bnab f + \frac{e}{m}\left(\textbf{E}+\bv\times \bB\right)\cdot \bnab_{\bv} f=0\,,
 \label{vlasov}\\
&&\textbf{E}= -\frac{n_i}{n_e}\textbf{u}_i\times \bB+\frac{1}{en_e}\textbf{J}\times \bB-\frac{\bnab P_e}{en_e}\,, \label{ohm}\\
&&\textbf{E}=-\nabla\Phi\,, 
\quad \bnab\times \bB=\mu_0\textbf{J}\,,\quad \bnab\cdot\bB=0\,, \label{maxwell}\\
&& \nabla\cdot\textbf{E}=e\left(n_i-n_e\right)\,, \quad n_i=\int d^3v\, f\,, \label{gauss}\\
&& P_e=n_e k_B T_{e0}\,. \label{eos}
\end{eqnarray}
Here, $f$ is the ion distribution function involved in the Vlasov equation (\ref{vlasov}), Eq. (\ref{ohm}) is the Ohm's law including the Hall and electron pressure-gradient terms,  $\sigma$ is the charge density given by:
\begin{eqnarray}
\sigma=e\left(n_i-n_e\right)=e\left(\int d^3v f -n_e\right)\,.
\end{eqnarray}
 The current density is
\begin{eqnarray}
\textbf{J}= \textbf{J}_i- en_e\textbf{u}_e\,,
\end{eqnarray}
where 
\begin{eqnarray}
\textbf{J}_i=en_i\textbf{u}_i=e\int d^3v\, \bv f\,, \label{u_J_ion}
\end{eqnarray}
is the kinetic ion current density; 
the  rest of the symbols are standard. To close the system we assume an isothermal electron fluid, i.e.,   $T_e=T_{e0}=\mbox{constant}$.

Let us consider the following normalization, which is  suitable in cases where the magnetic fields are strong and charge separation is insignificant:  
%introducing the following nondimensional quantities
\begin{eqnarray}
\label{normal1}
&&\tilde{r}=\frac{r}{\ell_i}\,,  \quad 
  \tilde{n}_j=\frac{n_j}{n_0}\, \   (j=e,i), \quad
  \tilde{f}=\frac{v_{A}^3 f}{n_0}\,, \quad 
  \tilde{P}_e=\frac{P_e}{m n_0 v_A^2} \,, \nl
&&\tilde{\bE}=\frac{\bE}{v_AB_0}\,, \quad 
  \tilde{\bB}=\frac{\bB}{ B_0}\,, \quad
  \tilde{\bJ}_i=\frac{\bJ_i}{en_0 v_{A}}\,, \quad  
  \tilde{\bv}=\frac{\bv}{v_{A}}\,, \quad
  \tilde{\bu}=\frac{\bu}{v_{A}}  
\end{eqnarray}
where $n_0$ and  $B_0$ are reference values for the density  and the magnetic field modulus and 
\begin{eqnarray}
\label{normal2}
\ell_i = v_A/\Omega\,,\quad v_A=\frac{B_0}{\sqrt{\mu_0 m n_0}}\,, \quad
\Omega &=& \frac{eB_0}{m}\,,
\end{eqnarray}
are the ion inertial length, the Alfv\'en velocity and the ion cyclotron frequency, respectively.

With this normalization, Eqs. (\ref{vlasov})--(\ref{gauss}) take the forms:
\begin{eqnarray}
&&\bv\cdot\bnab f+\left(\bE +\bv \times \bB\right)\cdot \bnab_{\bv} f=0\,, 
\label{vlasov_1}\\
&&-\nabla \Phi=\left[\frac{1}{n_e} (\bnab\times\bB-\bJ_i)\times 
\bB- \bnab \ln n_e^{\kappa} \right]\,,\label{ohm_1} \\
&&\bnab\times \bB =  \bJ\,,\quad 
\bnab\cdot\bB=0\,, \label{max_1}\\ 
&&\beta_{A}^2 \nabla \cdot \bE = (n_i-n_e)\,.\label{gauss_1}
\end{eqnarray} 
In the above equations, $\kappa=k_B T_{e0}/(mv_A^2)$ and  $\beta_A^2:= v_A^2/c^2$. By using this ordering scheme, in the  nonrelativistic limit  where  $\beta_A^2 \rightarrow 0$, the Gauss law (\ref{gauss_1}) implies the quasineutrality condition $n_i=n_e:=n$. Having restricted our study to the nonrelativistic regime, henceforth we will employ this condition instead of the Gauss equation.  In addition, since the hybrid model involves ion time scales, the electrons have enough time to reach thermal equilibrium. Accordingly, for the electronic fluid we will adopt the Boltzmann relation
\beq
\label{Boltz}
n_e=n_B\exp\left(\frac{\Phi}{\kappa}\right)\,,
\eeq
where $n_B$ is a free parameter.
 Using (\ref{Boltz}) in the Ohm's law (\ref{ohm_1}) it turns out that $\left(\bnab\times \bB-\bJ_i\right)\times \bB=0$, which implies that the vector $\bnab\times \bB-\bJ_i$ is parallel to $\bB$; therefore, we can write
 \beq
 \label{current}
 \bnab\times \bB=\lambda \bB +\bJ_i\,,
 \eeq
 where in general $\lambda$ can be a spatially dependent function  $\lambda=\lambda(\textbf{r})$. In the absence of the ion current density $\textbf{J}_i$, this would correspond to non-linear force-free magnetic fields that seem to describe best the observed MFRs with peaked on-axis axial field \cite{Yang2014}. For simplicity though, here  we will employ $\lambda$ as a free parameter. According to (\ref{current}),  the current density splits into a Beltrami electron-fluid component, $\lambda \bB$,  and a kinetic ion component, 
 $\bJ_i= \int d^3v f \bv$. Summarizing up to this point, the equilibrium equations have been reduced to (\ref{current}), the Vlasov equation (\ref{vlasov_1}) for the ions together with the quasineutrality condition and the Boltzmann equation (\ref{Boltz}) for the electrons.

Furthermore, we will pursue constructing one-dimensional steady states of a cylindrical plasma with radius $r_s$ of the circular cross-section. The respective dimensionless radius is $\tilde{r}_s=r_s/l_i=:r_0$.  In cylindrical coordinates $(r,\phi,z)$ the equilibrium quantities depend only on $r$. The magnetic field has both axial and azimuthal components  (screw-pinch) including a constant  axial background magnetic field $B_g \hat{e}_z $.   The kinetic ions have three constant of motion, i.e. the energy $H=v^2/2 +\Phi$ and the canonical momenta conjugate to ignorable coordinates $\phi$ and $z$:
\beq
\label{momenta}
p_\phi=\frac{B_g r^2}{2}+r v_\phi + r A_\phi(r)\,, \quad p_z=v_z + A_z(r)\,.
\eeq
It is recalled that we use dimensionless quantities; $\bA=\left(r B_g/2 + A_\phi(r)\right)\hat{e}_\phi + A_z(r) \hat{e}_z$ 
%($\tilde{\bA}=\bA/(v_A B_0/\Omega)$) 
is the vector potential with the magnetic field, $\bB=\bnab \times \bA$, given by
\beq
\label{mf}
\bB=B_\phi(r)\hat{e}_\phi + \left(B_g + B_z(r)\right) \hat{e}_z\,.
\eeq
According to the Jean's theorem any function of $H$, $p_\phi$ and $p_z$  is solution of the Vlasov equation (\ref{vlasov_1}). Here we will adopt the ion distribution function introduced in \cite{Ng} and employed in \cite{KaMo}:
\beq
\label{df}
f(H,p_\phi,p_z)=2^{-1/2}\pi^{-3/2}e^{-H}\left[1-d_1 \exp\left(-d_2 p_\phi^2-d_3 p_z^2\right)\right]\,, 
\eeq
where $d_1$, $d_2$ and $d_3$ are free parameters with $d_1<1$ so that $f$ is always positive. For $d_1=0$ (\ref{df}) becomes Maxwellian and the ion current density as well the macroscopic ion velocity vanish; thus, the equilibrium becomes static. Setting $rA_\phi(r)=:\Psi(r)$, Eq. (\ref{current}) is equivalent to the following set of first-order, quasilinear ODEs for the functions $
\Psi(r)$, $A_z(r)$, $B_\phi(r)$ and $B_z(r)$:

\beqa
B_\phi=-\dadb{A_z}{r} \label{Eq_1} \,, \\
B_z=\frac{1}{r}\dadb{\Psi}{r} \,,  \label{Eq_2}  \\
-\dadb{B_z}{r}=\lambda B_\phi +J_{i\phi} \,, \label{Eq_3} \\
\frac{1}{r}\dadb{(r B_\phi)}{r}=\lambda\left(B_g +B_z\right) + J_{iz} \label{Eq_4}\,.
\eeqa
Here, the ion-current-density components are given in terms of $f$ by
\beq
\label{cur_comp}
J_{ij}=\int d^3v\, v_j f\,, \quad j=\phi,z 
\eeq
and the electrostatic potential, appearing in the ion energy, can be expressed in terms of $\Psi$ and $A_z$ by means of the quasineutrality condition;  with the aid of  (\ref{Boltz}), (\ref{df}) and the second of (\ref{gauss}) we find
\beqa
\label{Phi}
\Phi(r)&=&\frac{\kappa}{\kappa + 1}\log\left(\frac{K_1(r)}{n_B}\right)\,, \label{Phi_1} \\
K_1(r)&=&2^{-1/2}\pi^{-3/2}\int d^3 v\, \exp\left(-\frac{v^2}{2}\right)\left[1-d_1 \exp\left(-d_2 p_\phi^2-d_3 p_z^2\right)\right]. \label{Phi_2}
\eeqa
The equilibrium can be completely constructed by solutions of the respective initial-value problem.

%%%%%%%%%%%%%%%%%%%%%%%%%%%%%%%%%%%%%%%%%%%%%%%%%%%%%%%%%%%%%%%%%%%%%%%

\section{Equilibrium solutions and parametric impact}\

In association with the set of ODEs (\ref{Eq_1})-(\ref{Eq_4}) we will impose the following initial conditions\footnote{It may be noted that the term ``initial conditions" here is used in mathematical setting on the understanding that the independent variable, $r$, is spatial.}: 
\beq
\label{conditions}
\Psi(10^{-10})=10^{-10}\,, \quad  A_z(10^{-10})=0.5\,, \quad B_\phi(10^{-10})=10^{-10}\,, \quad  B_z(r_0)=0 \,.
\eeq
The fourth of the above conditions guarantees that the axial magnetic field on the plasma surface, $r=r_0$, becomes equal to the background field $B_g$. It may be noted that for a related initial value problem involving the same initial value for all the unknown functions,  the  Picard-Lindel\"of theorem provides a sufficient condition for the existence and uniqueness of solution. We chose the following values for the reference quantities (relations (\ref{normal1}) and (\ref{normal2})): $n_0=10^{19}$  m$^{-3}$ and $B_0=1$  T resulting in $\Omega=9.58\times 10^7$  rad/sec, $v_A=6.9\times 10^{6}$ m/sec and $li=0.072$ m. The parameters $T_{e0}$, $n_B$, $B_g$, $\lambda$, $r_s$,  $d_1$, $d_2$ and $d_3$ are free. The impact of $T_{e0}$ on the solutions, through $\kappa=k_B T_{e0}/(mv_A^2)$ in (\ref{Phi_1}) is negligible. It only affects the electron thermal pressure (\ref{eos}). For this reason we fixed its value to $T_{e0}=10^{8}$  $^o$K . Also, very weak is the impact of $n_B$ and we fixed its value to 10. 

\begin{figure}[h]
%\vspace{-0.4cm}
\begin{center}
\includegraphics[width=0.49\linewidth]{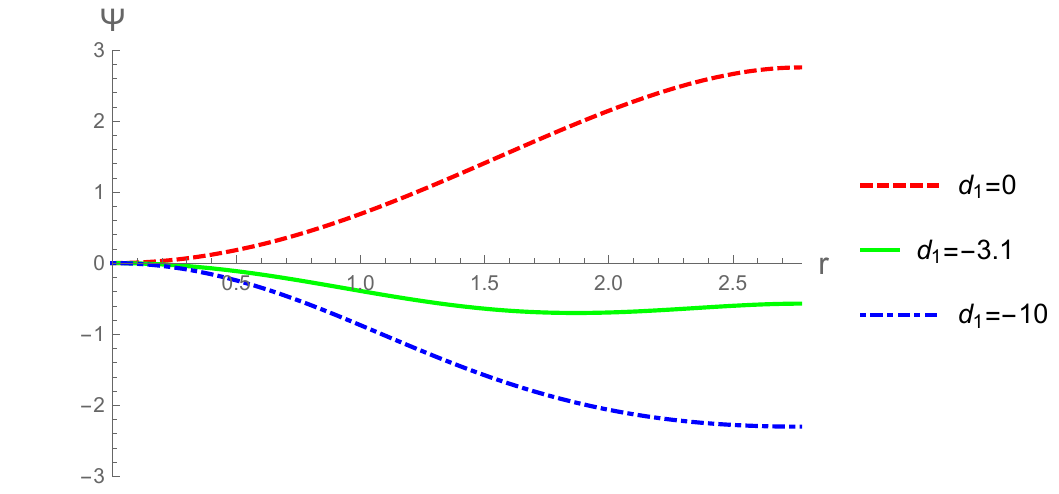}
\includegraphics[width=0.49\linewidth]{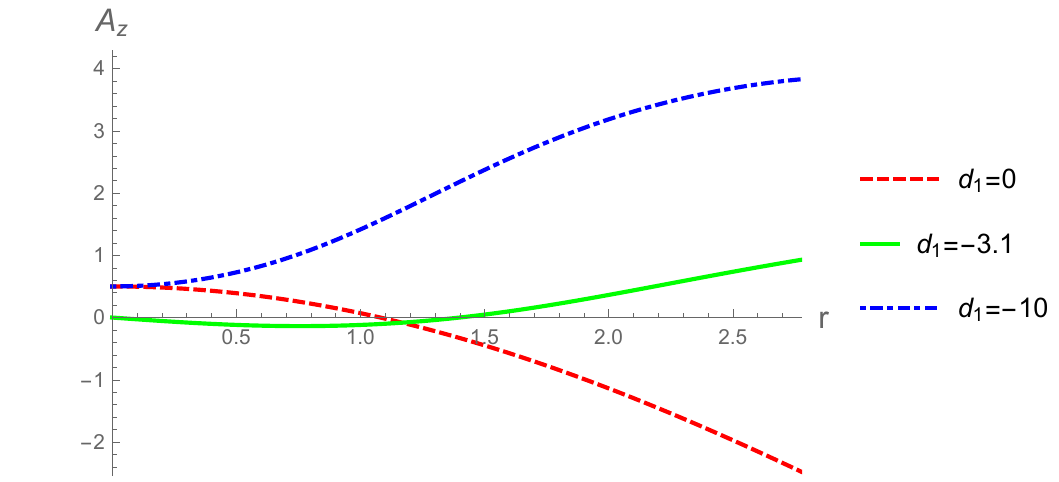}
\includegraphics[width=0.49\linewidth]{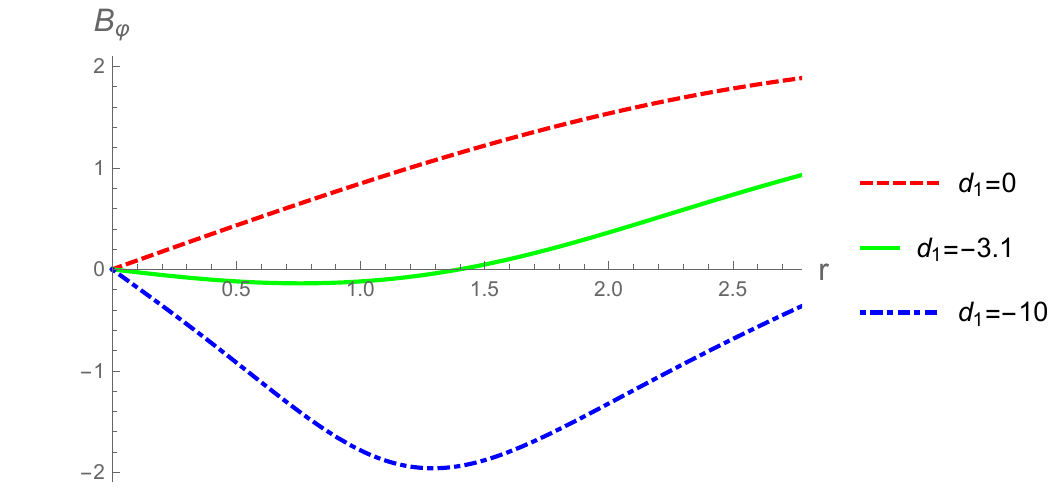}
\includegraphics[width=0.49\linewidth]{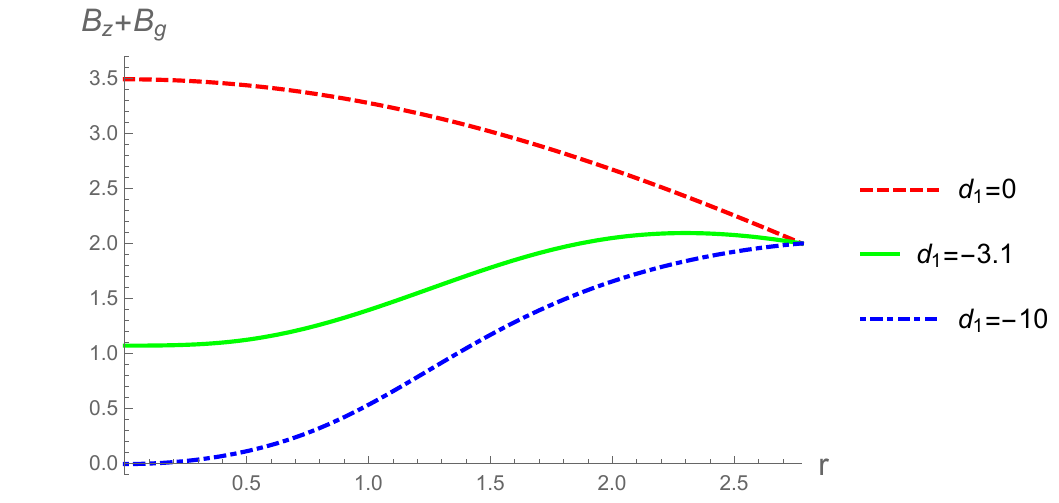} 
\end{center}
\caption{Numerical solutions of the set of ODEs (\ref{Eq_1})-(\ref{Eq_4}) under the initial conditions (\ref{conditions}) for three different values of the parameter $d_1$ contained in the ion distribution function (\ref{df}). The value $d_1=0$ corresponds to a Maxwellian distribution function. The values of the other free parameters are  $d_2=d_3=1$, $\lambda=0.5$, $B_g=2$,   and $r_s=0.2\ m$ ($r_0=r_s/l_i=2.78$). The down-right plot of the axial magnetic field includes the background field $B_g$. The value $d_1=-3.1$ is the largest negative for which the ion-pressure-tensor element, $P_{\phi z}$, is throughout the plasma non-negative.}
                                                          \label{d_1}
\end{figure}

\begin{figure}[h]
%\vspace{-0.4cm}
\begin{center}
\includegraphics[width=0.49\linewidth]{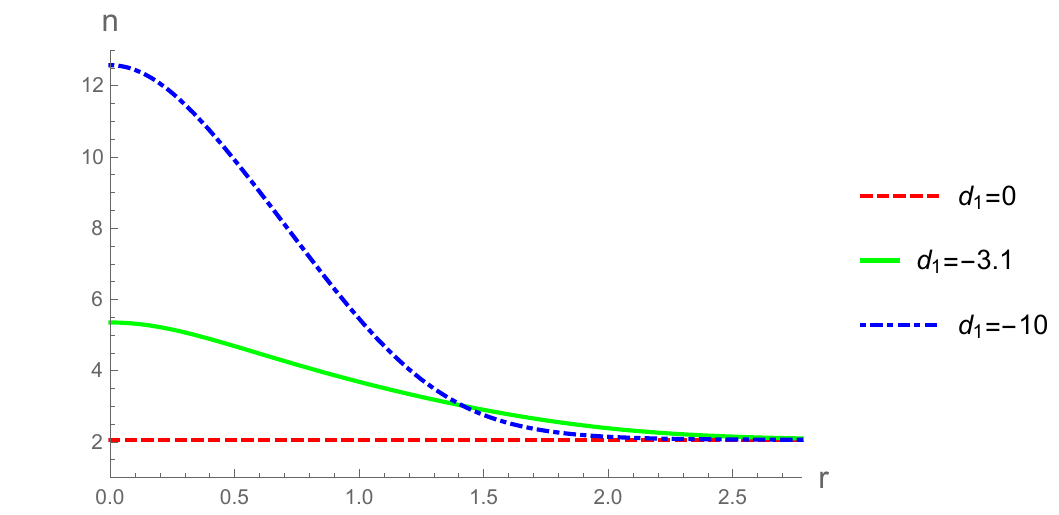}
\includegraphics[width=0.49\linewidth]{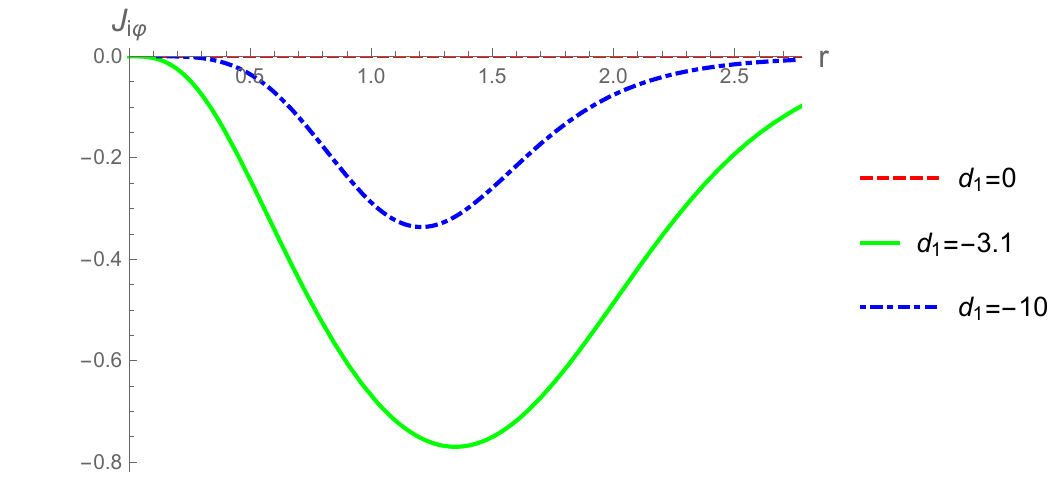}
\includegraphics[width=0.49\linewidth]{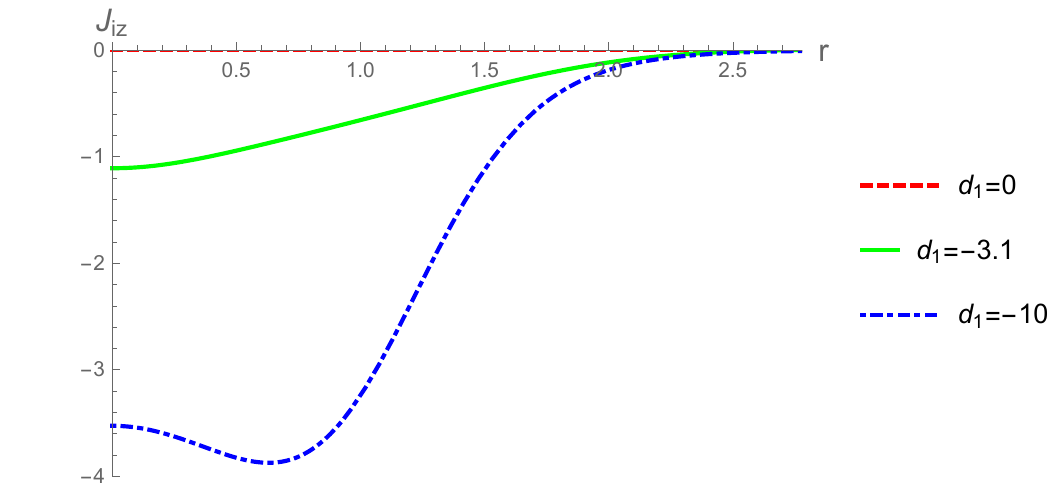}
\includegraphics[width=0.49\linewidth]{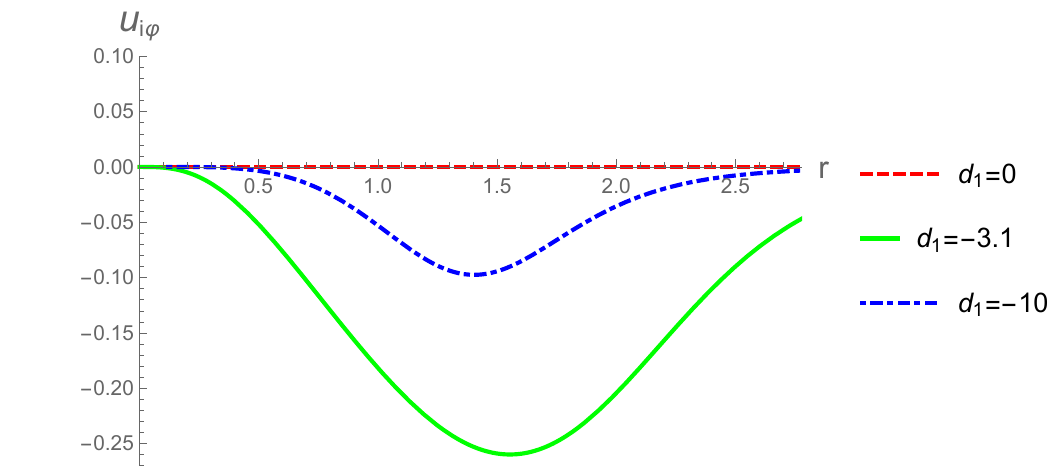}
\includegraphics[width=0.49\linewidth]{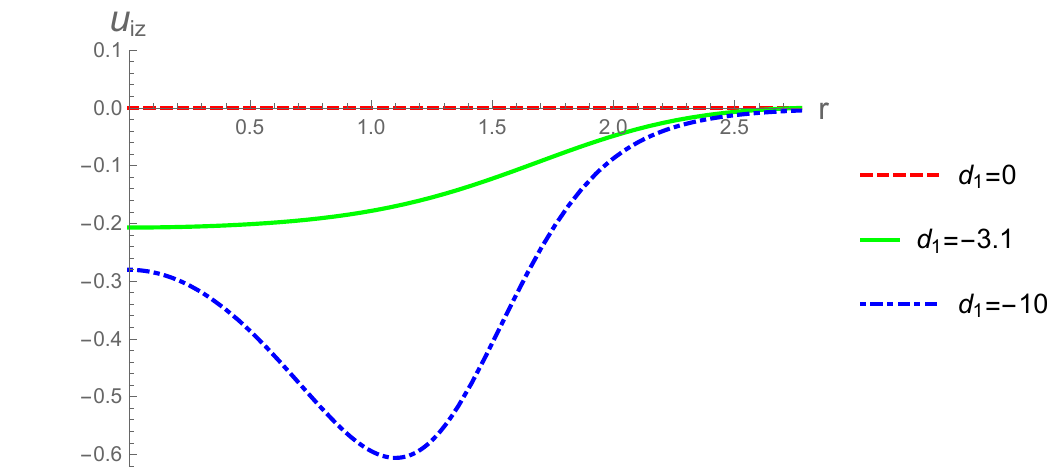}
\includegraphics[width=0.49\linewidth]{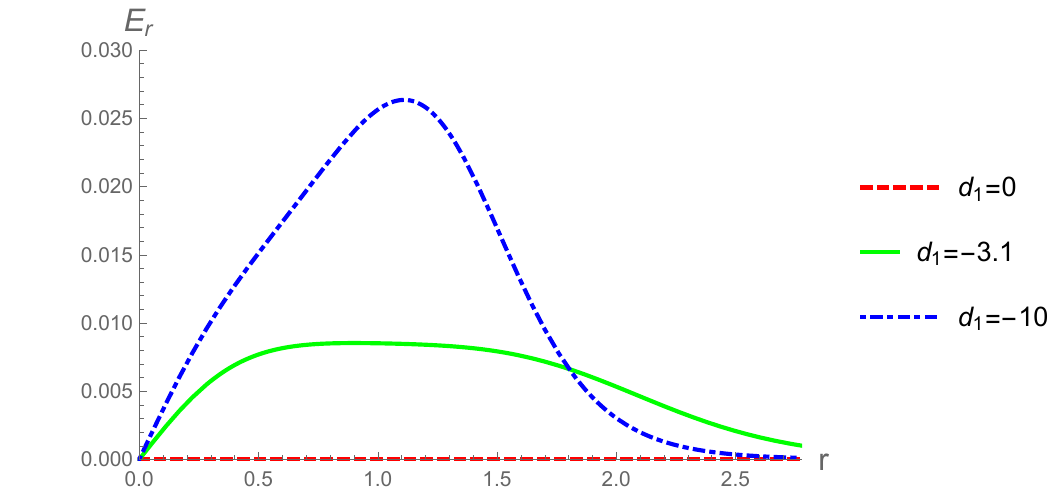}
\includegraphics[width=0.49\linewidth]{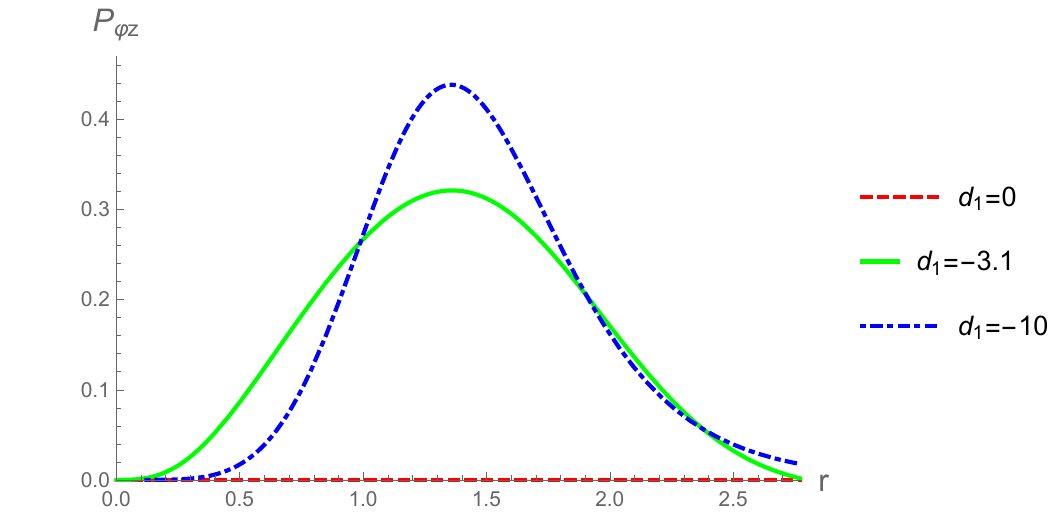}
\includegraphics[width=0.49\linewidth]{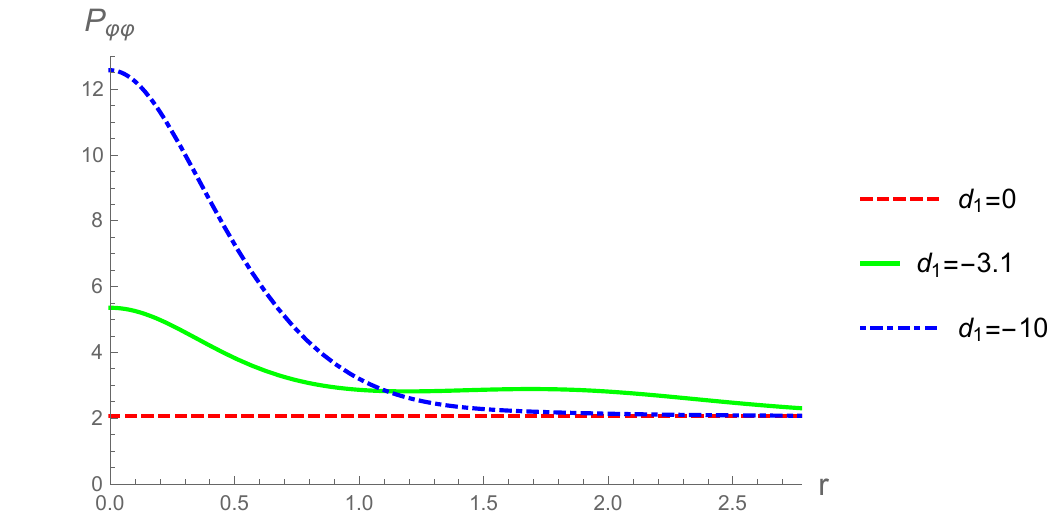}
\includegraphics[width=0.49\linewidth]{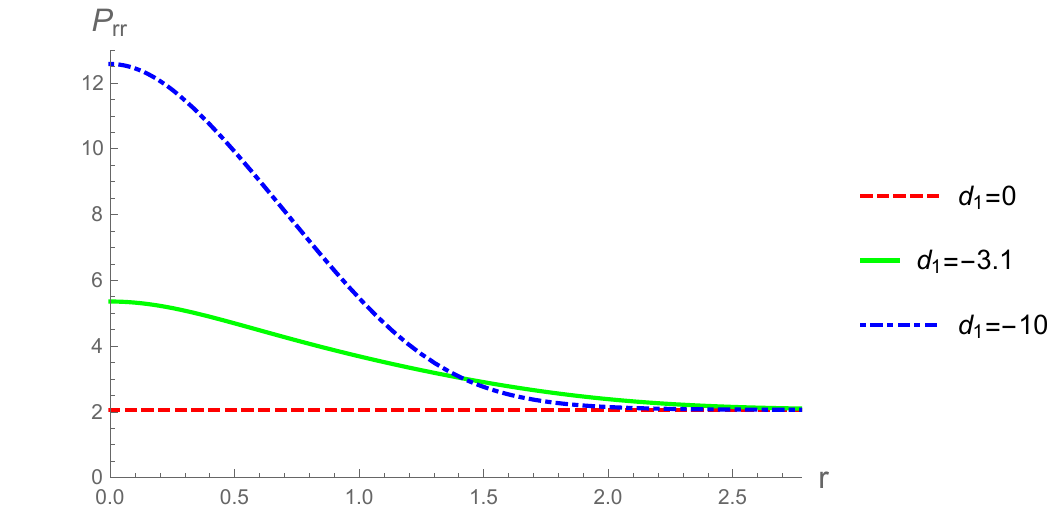}
\includegraphics[width=0.49\linewidth]{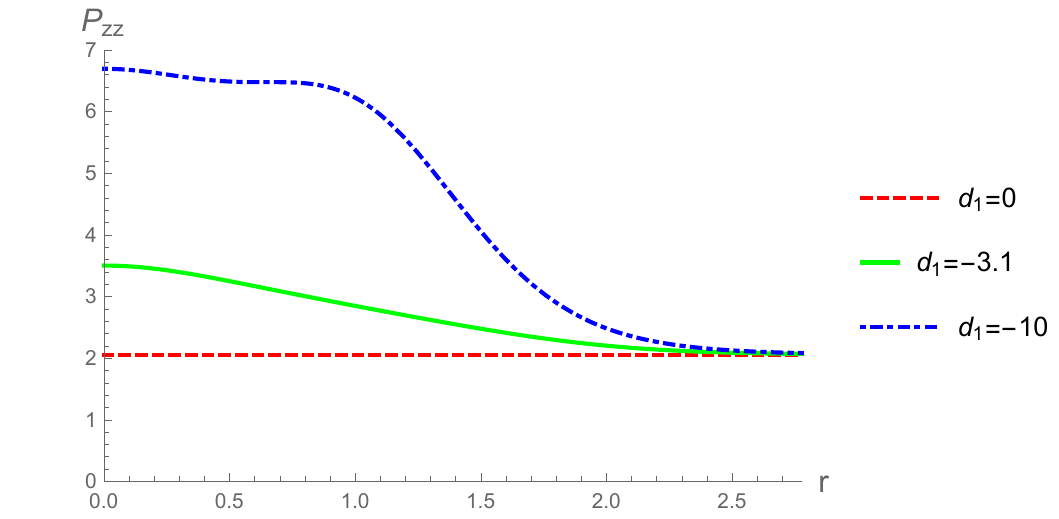}
\end{center}
\caption{Radial profiles of the density, $n$, the components of the ion current density, $\bJ_i$, and of the macroscopic ion velocity, $\bu_i=\bJ_i/n$, the radial electric field, $E_r$, and the elements $P_{jk}\,, \ (j,k)=(r,\, \phi,\,  z )$ of the ion pressure tensor for the solutions of Fig. \ref{d_1}. The profiles for  $d_1=0$ correspond to a Maxwellian ion distribution function. }
                                                          \label{d_1_profiles}
\end{figure}

\begin{figure}[h]
%\vspace{-0.4cm}
\begin{center}
\includegraphics[width=0.49\linewidth]{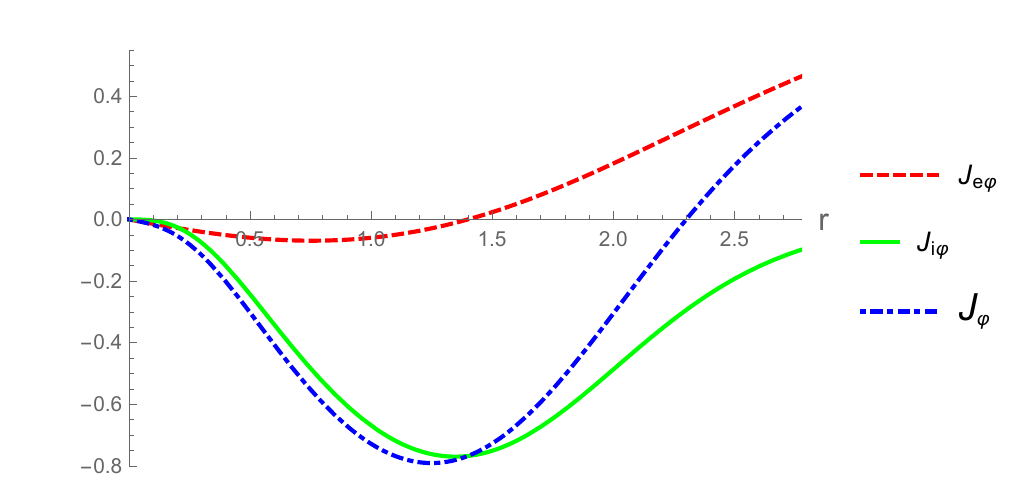}
\includegraphics[width=0.49\linewidth]{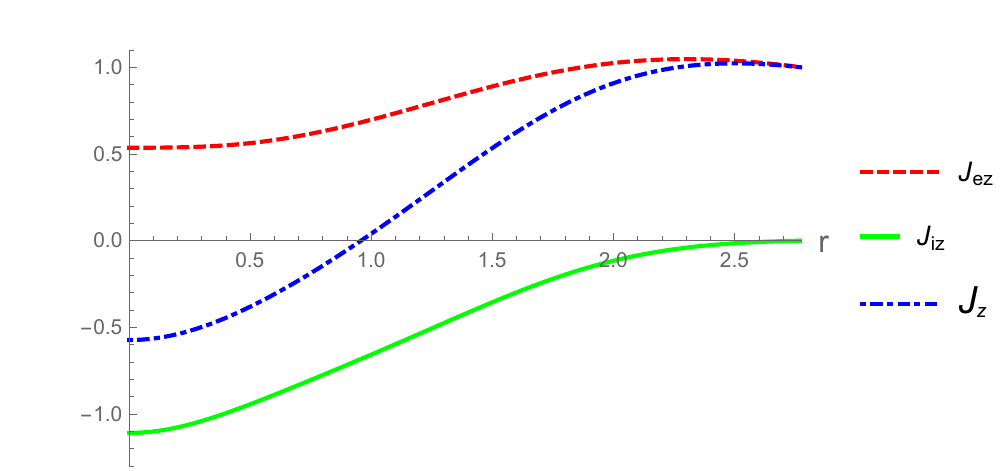}
\end{center}
\caption{Radial profiles of the electron, ion and total current density for the solution of Fig. \ref{d_1} with $d_1=-3.1$.  }
                                                          \label{currents}
\end{figure}

\begin{figure}[h]
%\vspace{-0.4cm}
\begin{center}
\includegraphics[width=0.49\linewidth]{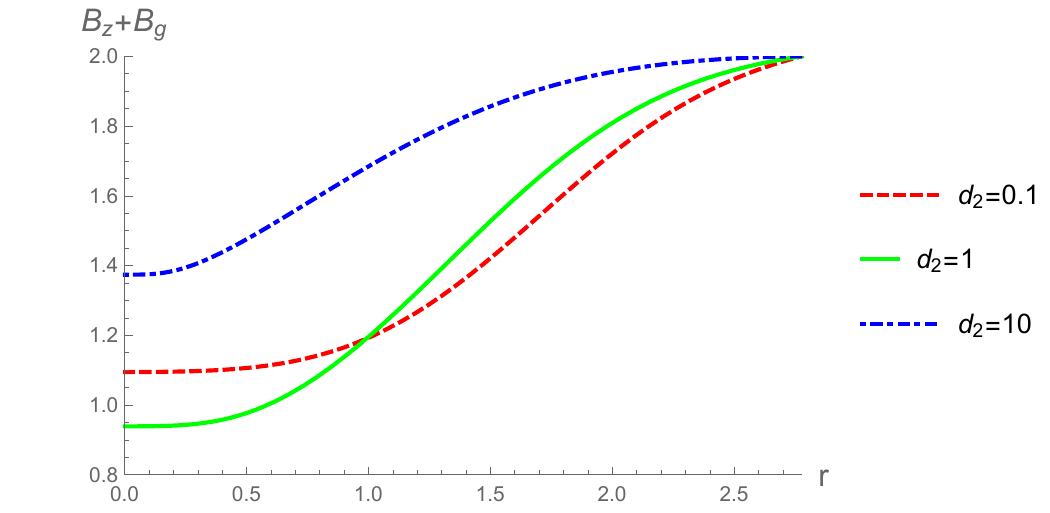}
\includegraphics[width=0.49\linewidth]{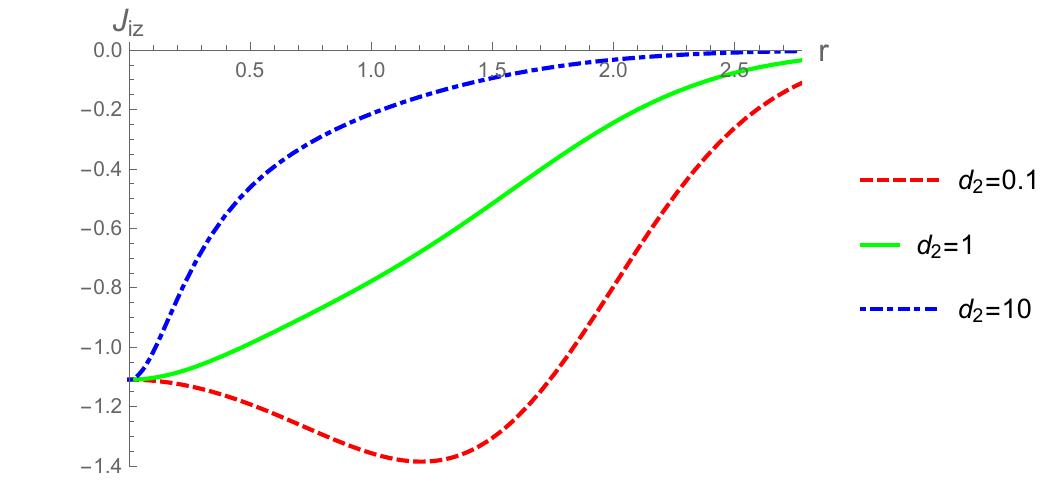}
\includegraphics[width=0.49\linewidth]{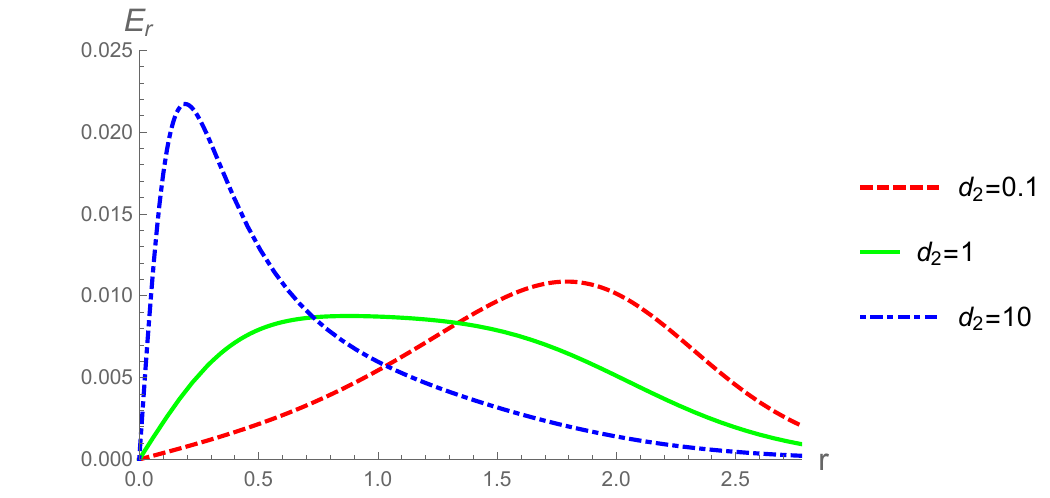}
\includegraphics[width=0.49\linewidth]{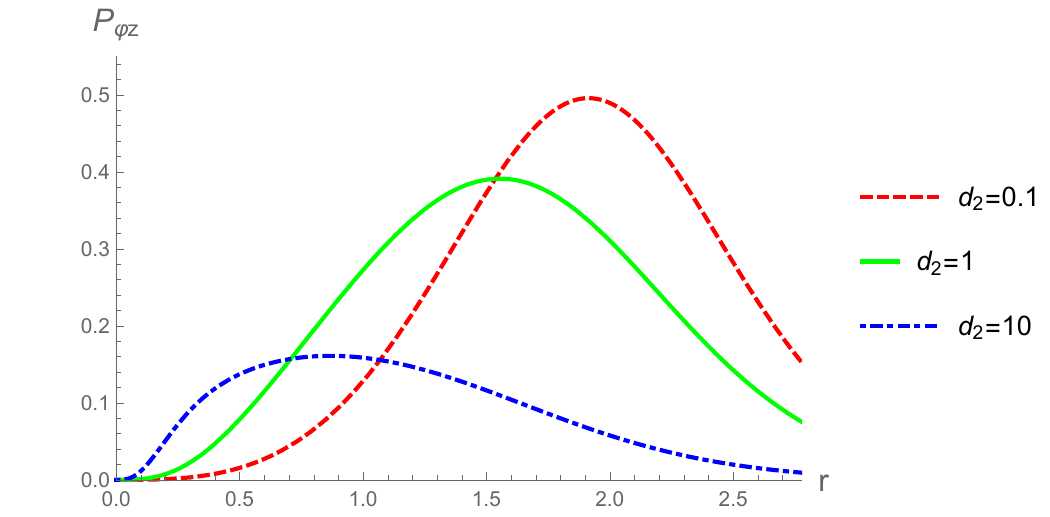}
\end{center}
\caption{Impact of the parameter $d_2$ contained in the ion distribution function (\ref{df}) on the total axial magnetic field, $B_z+B_g$, the axial ion current density, $J_{iz}$, the radial electric field, $E_r$,  and the non-gyrotropic element of the ion pressure tensor, $P_{\phi z}$. The values of the other free parameters are $d_1=-3.1$, $d_3=1$, $\lambda=0.1$,  $B_g=2$,   and $r_0=2.78$.   }
                                                          \label{d_2}
\end{figure}

\begin{figure}[h]
%\vspace{-0.4cm}
\begin{center}
\includegraphics[width=0.49\linewidth]{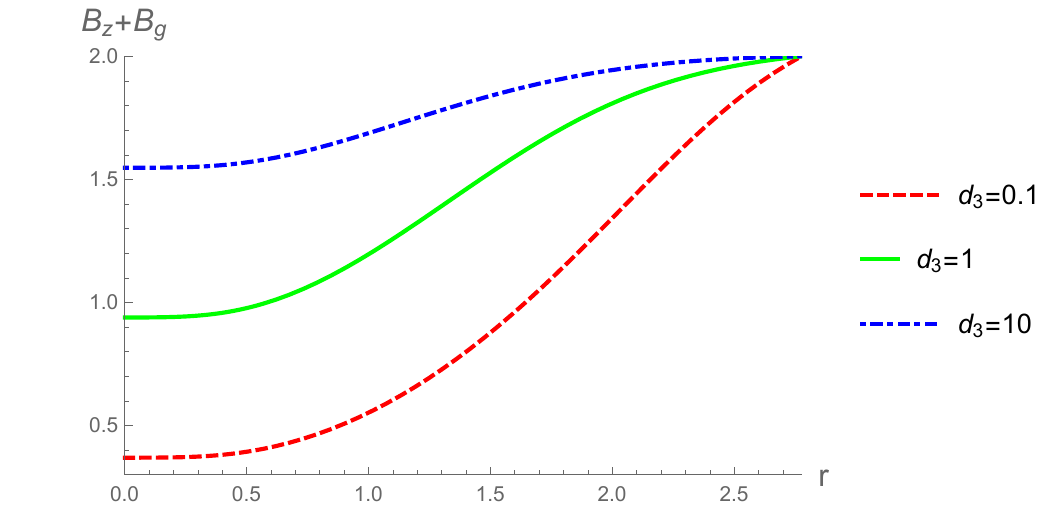}
\includegraphics[width=0.49\linewidth]{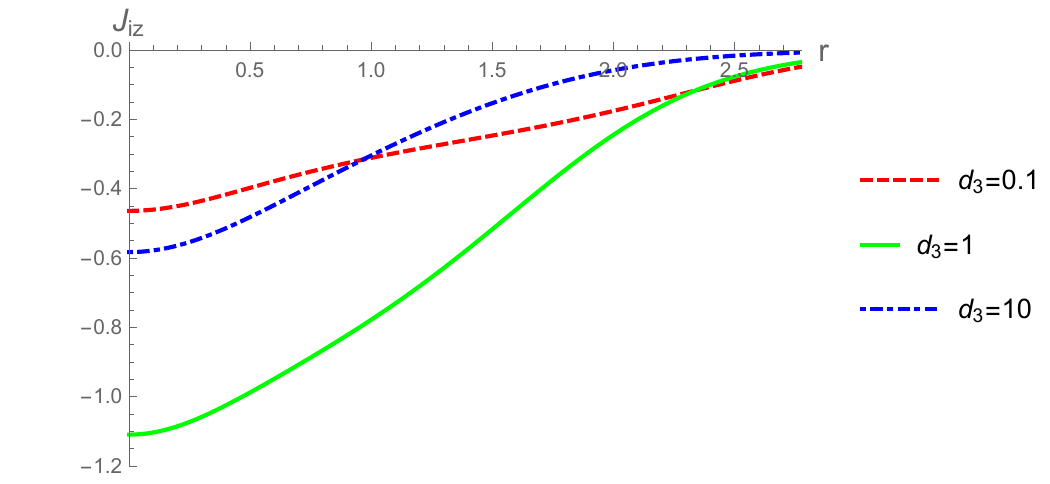}
\includegraphics[width=0.49\linewidth]{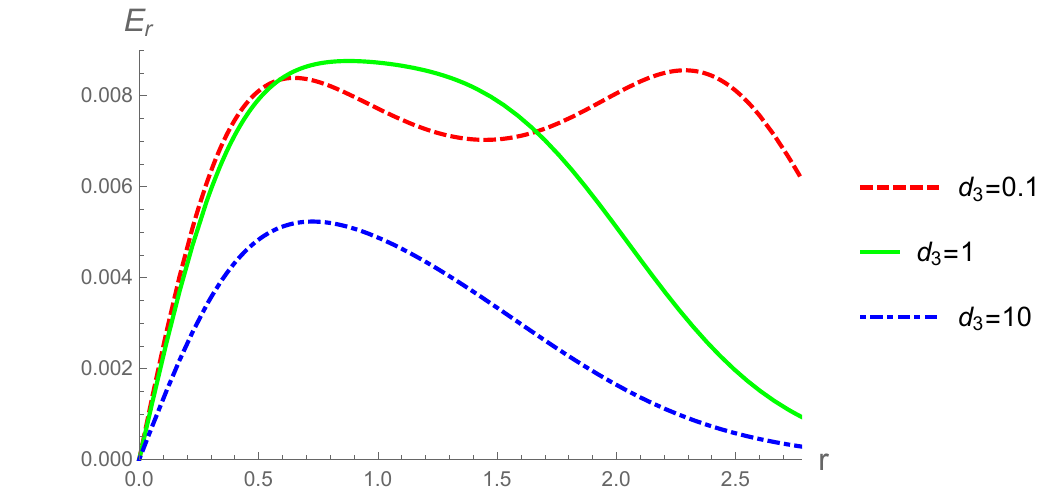}
\includegraphics[width=0.49\linewidth]{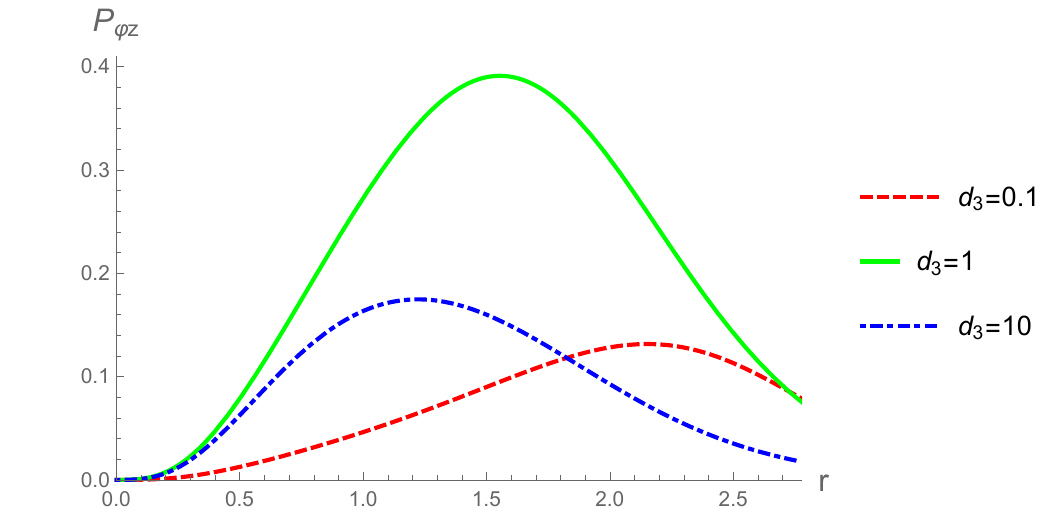}
\end{center}
\caption{Impact of the parameter $d_3$ contained in the ion distribution function (\ref{df}) on the total axial magnetic field, $B_z+B_g$, the axial ion current density, $J_{iz}$, the radial electric field, $E_r$,  and the non-gyrotropic element of the ion pressure tensor, $P_{\phi z}$. The values of the other free parameters are $d_1=-3.1$, $d_2=1$, $\lambda=0.1$, $B_g=2$,   and $r_0=2.78$.   }
                                                          \label{d_3}
\end{figure}

\begin{figure}[h]
%\vspace{-0.4cm}
\begin{center}
\includegraphics[width=0.49\linewidth]{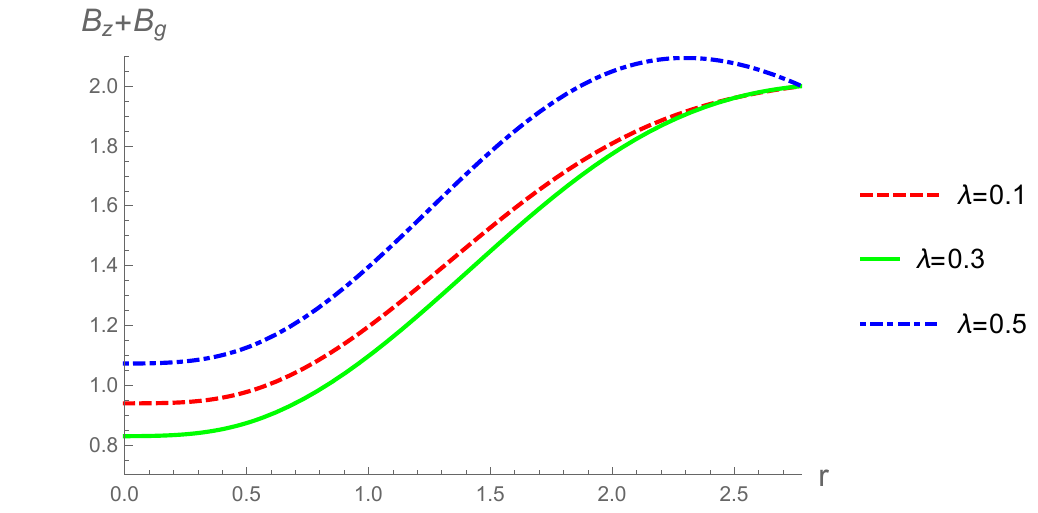}
\includegraphics[width=0.49\linewidth]{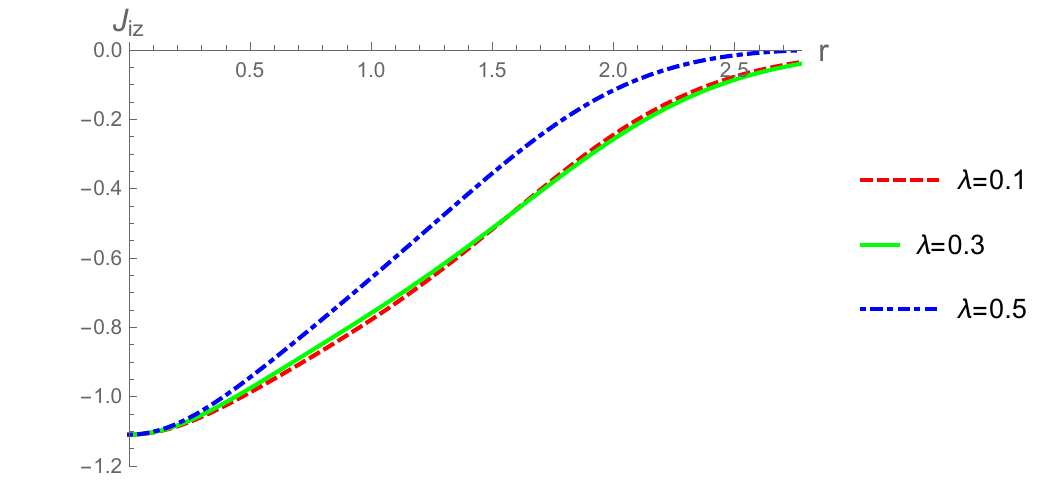}
\includegraphics[width=0.49\linewidth]{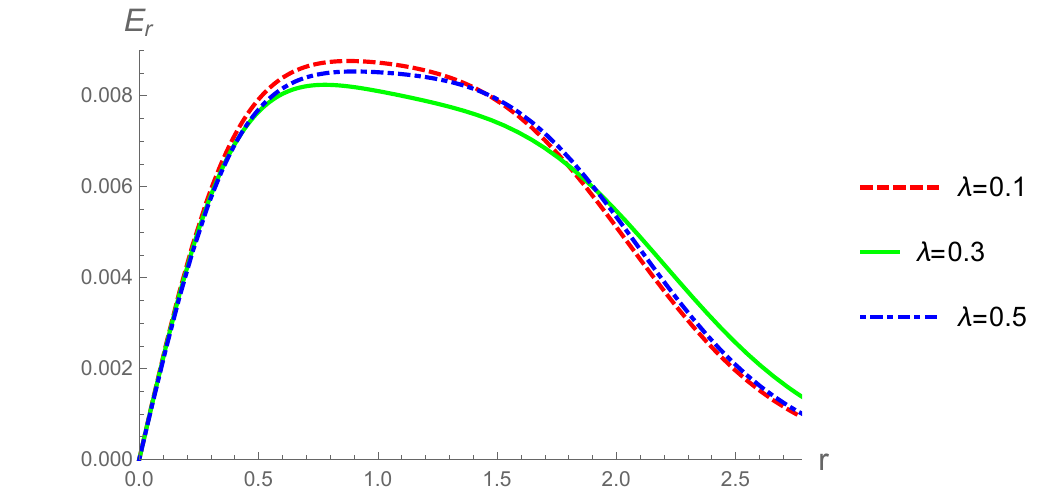}
\includegraphics[width=0.49\linewidth]{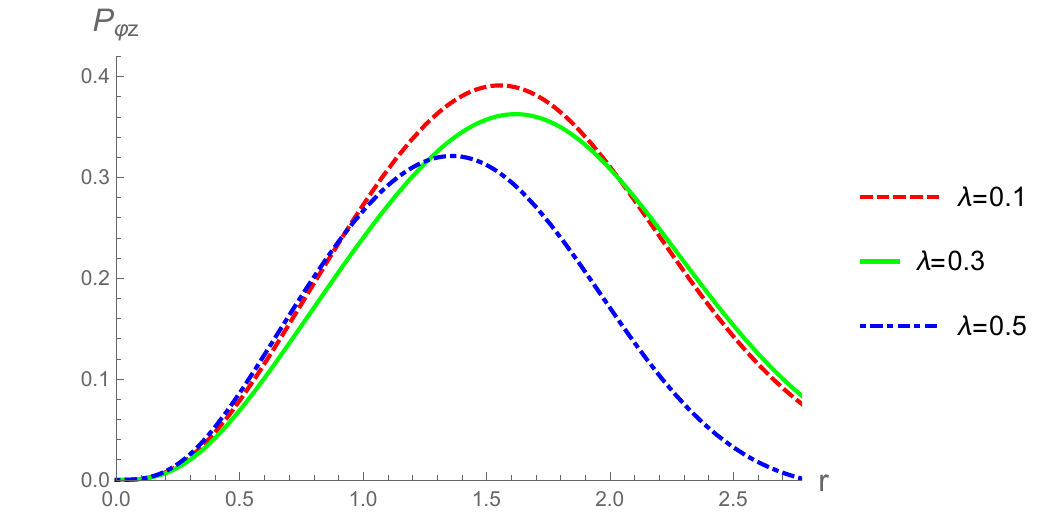}
\end{center}
\caption{Impact of the Beltrami parameter $\lambda$ involved in Eq. (\ref{current}) on the total axial magnetic field, $B_z+B_g$, the axial ion current density, $J_{iz}$, the radial electric field, $E_r$,  and the non-gyrotropic element of the ion pressure tensor, $P_{\phi z}$. The values of the other free parameters are $d_1=-3.1$, $d_2=d_3=1$, $B_g=2$,   and $r_0=2.78$.   }
                                                          \label{lambda}
\end{figure}

\begin{figure}[h]
%\vspace{-0.4cm}
\begin{center}
\includegraphics[width=0.49\linewidth]{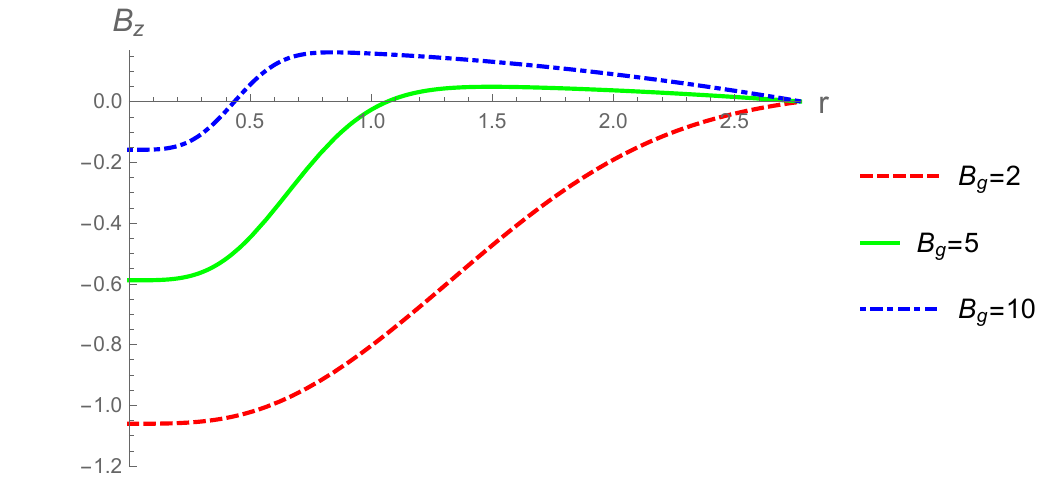}
\includegraphics[width=0.49\linewidth]{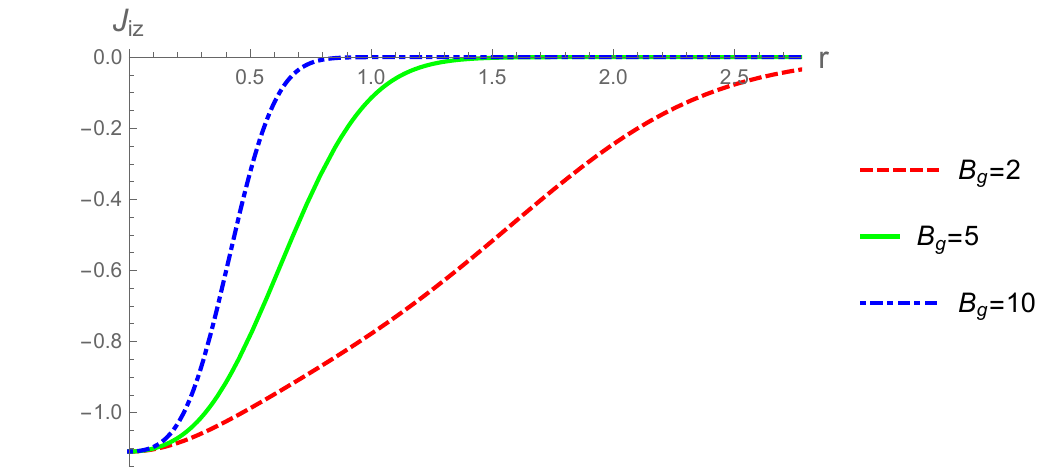}
\includegraphics[width=0.49\linewidth]{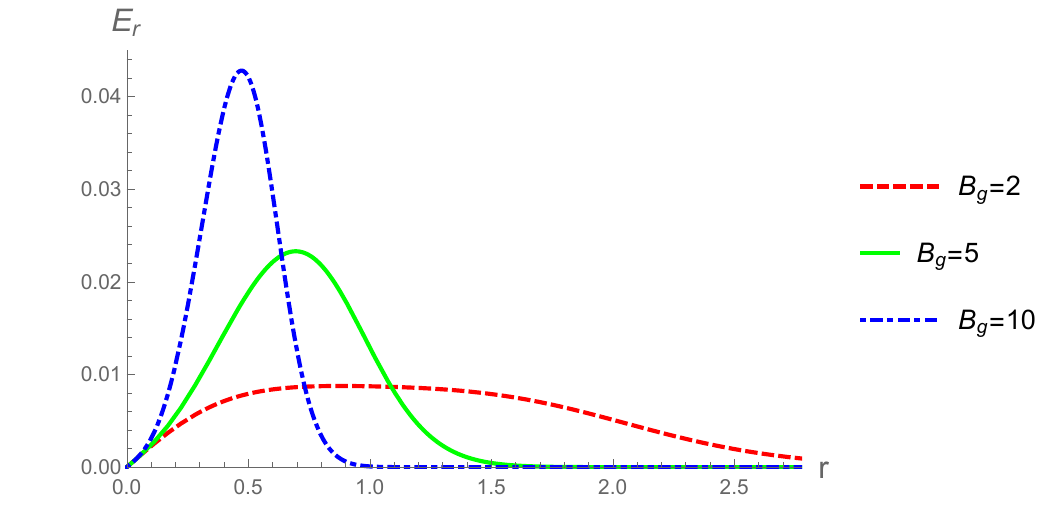}
\includegraphics[width=0.49\linewidth]{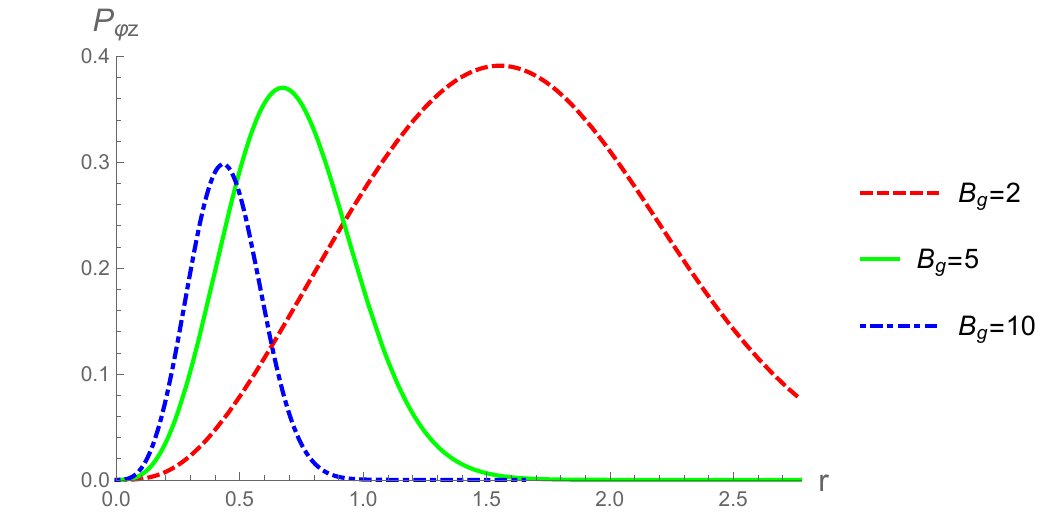}
\end{center}
\caption{Impact of the background axial magnetic field $B_g$ on the axial magnetic field, $B_z$, produced by the azimuthal plasma current,  the axial ion current density, $J_{iz}$, the radial electric field, $E_r$,  and the non-gyrotropic element of the ion pressure tensor, $P_{\phi z}$. The values of the other free parameters are $d_1=-3.1$, $d_2=d_3=1$,  $\lambda=0.1$  and $r_0=2.78$.   }
                                                          \label{B_g}
\end{figure}

\begin{figure}[h]
%\vspace{-0.4cm}
\begin{center}
\includegraphics[width=0.49\linewidth]{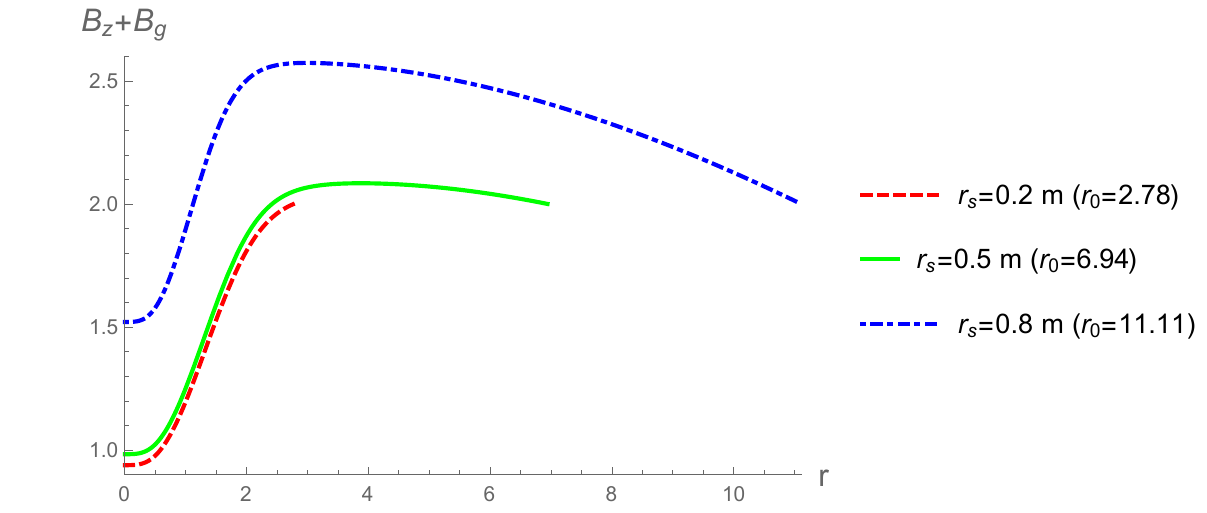}
\includegraphics[width=0.49\linewidth]{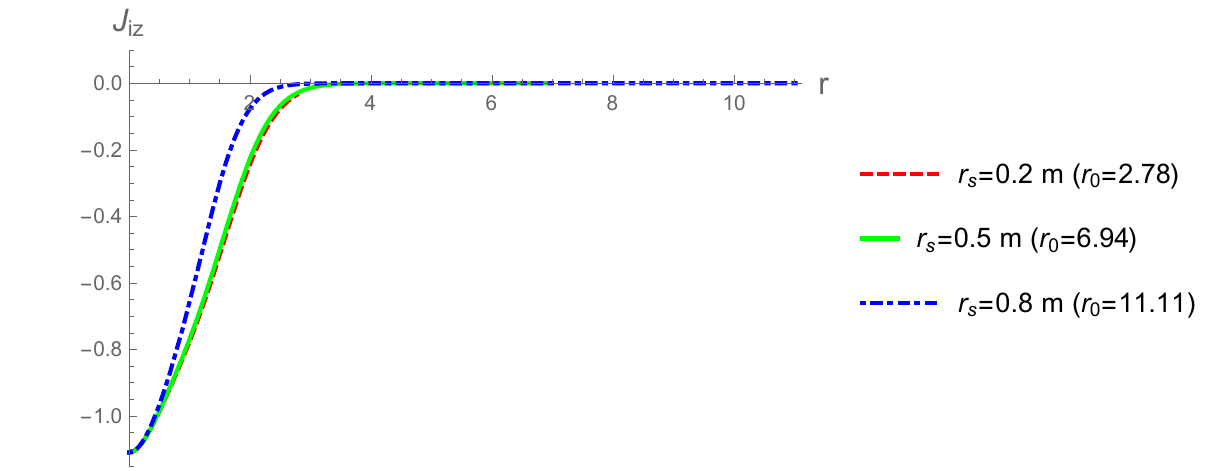}
\includegraphics[width=0.49\linewidth]{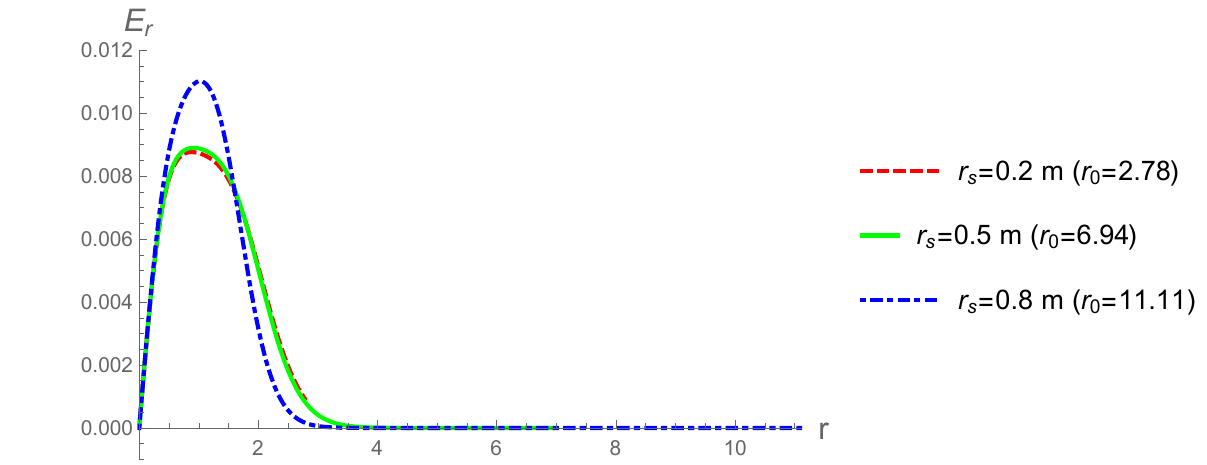}
\includegraphics[width=0.49\linewidth]{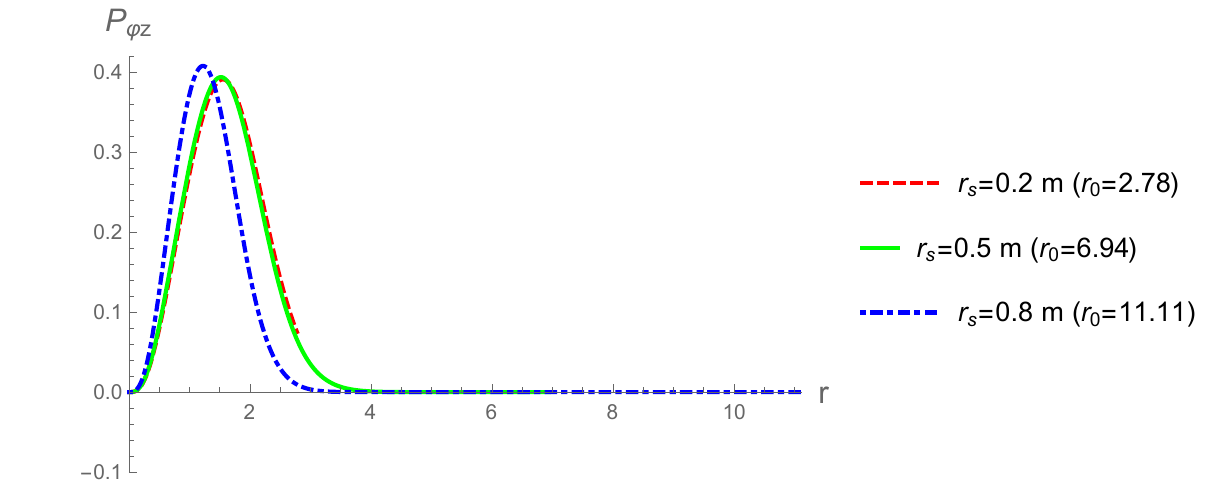}
\end{center}
\caption{Impact of the radius of the cylindrical plasma column  on the total axial magnetic field, $B_z+B_g$, the axial ion current density, $J_{iz}$, the radial electric field, $E_r$,  and the non-gyrotropic element of the ion pressure tensor, $P_{\phi z}$. The values of the other free parameters are $d_1=-3.1$, $d_2=d_3=1$, $\lambda=0.1$,   and $B_g=2$.   }
                                                          \label{r_s}
\end{figure}

After performing the integrations involved in the velocity space analytically we solved the initial value problem of the ODEs (\ref{Eq_1})-(\ref{Eq_4}) with the initial conditions  (\ref{conditions}) for different values of the other free parameters.  Solutions for three  values of the parameter $d_1$ is given in Fig. \ref{d_1}.
For each solution the equilibrium can be completely constructed including the elements of  the ion pressure tensor
\beq
\label{P_ion}
P_{jk} = \int\, d^3v\, v_j v_k f\,, \quad (j,k)=(r,\, \phi,\, z)\,.
\eeq
This tensor is symmetric, non-gyrotropic with $P_{r\phi }=P_{rz}=0$ but $P_{\phi z}\neq 0 $.
Radial profiles of the various equilibrium quantities for the solutions  of Fig. \ref{d_1} are given in Fig. \ref{d_1_profiles}. As expected, the profiles of the components of the macroscopic ion velocity, $\bu_i=\bJ_i/n$ are similar in shape with those of the ion current density, $\bJ_i$.
Also, electron, ion and total current density profiles for the solution of Fig. \ref{d_1} with $d_1=-3.1$ are given in Fig. \ref{currents}.
In addition, we examined the impact of the parameters  $d_2$, $d_3$, $\lambda$, $B_g$ and $r_s$ on the equilibrium. In each case respective radial profiles of the quantities $B_z+B_g$ (or $B_z$),  $J_{iz}$, $E_r$ and $P_{\phi z}$  are given in Figs. \ref{d_2}, \ref{d_3}, \ref{lambda}, \ref{B_g} and \ref{r_s}. 

The above calculations led to the following conclusions:
\begin{enumerate}
\item All the free parameters affect significantly the magnetic properties of the plasma resulting in a variety of equilibria either diamagnetic, or paramagnetic, or mixed associated with non monotonic variation of $B_z(r)$ as can be seen in Figs. \ref{d_1}, \ref{d_2}, \ref{d_3}, \ref{lambda}, \ref{B_g}, and \ref{r_s}. For example, in Fig. \ref{d_1} for $d_1=-10$  the equilibrium is diamagnetic, it is mixed for $d_1=-3.1$ with a broad  diamagnetic inner region and a slightly paramagnetic outer region, and it becomes paramagnetic for $d_1=0$. It is recalled that for $d_1=0$ the ion distribution function (\ref{df}) becomes Maxwellian. Similar is the impact of $B_g$ in Fig. \ref{B_g} and of $r_s$ in Fig. \ref{r_s}, where  in the mixed cases the outer paramagnetic regions (red dashed-dotted curves) are broader and more pronounced.  
\item The axial ion current density profiles can be either peaked on-axis (Fig. \ref{d_1_profiles} for $d_1=-3.1$; Fig. \ref{d_2} for $d_2=0.1$ and $d_2=1$; Figs. \ref{d_3}, \ref{lambda}, \ref{B_g}, \ref{r_s}) or peaked off-axis (Fig. \ref{d_1_profiles} for $d_1=-10$; Fig. \ref{d_2} for $d_2=10$).  The  peaked profiles can be either convex (Fig. \ref{d_2} for $d_2=1$) or concave (Fig. \ref{d_2} for $d_2=10$). Also, they   become more centrally localized for larger values of $B_g$ (Fig. \ref{B_g}) and remain centrally localized for larger values of $r_s$ (Fig. \ref{r_s}).
\item For $d_1<0$ the density, $n$, is peaked on axis (Fig. \ref{d_1_profiles}). Since the electron fluid temperature is constant,  similar to the density profiles  are the thermal electron pressure profiles (Eq. (\ref{eos})). For a Maxwellian ion distribution function ($d_1=0$) the density becomes constant and the ion pressure becomes scalar, $P_{rr}=P_{\phi \phi}=P_{zz}$= constant and $P_{\phi z}=0$ (cf. Fig. \ref{d_1_profiles}) and therefore the total pressure becomes constant.
\item In certain case the results show a similarity  in location or/and shape between the radial electric field, $E_r$, the non gyrotropic element of the ion pressure tensor,  $P_{\phi z}$,  and the components of the macroscopic sheared ion velocity, $\bu_i$   (Figs. \ref{d_1_profiles}, \ref{d_2}, \ref{lambda}, \ref{B_g}, and \ref{r_s}), thus indicating a potential correlation one another of these quantities.  It is noted that there is  experimental \cite{CaPu,BiRa} and theoretical \cite{TyCz,LiWa,YaWi} evidence in the improved confinement regimes of laboratory fusion plasmas of an interplay between sheared zonal flows, radial electric fields, Reynolds or residual pressure, symmetry breaking, turbulence and transport regulation.
\end{enumerate}

\section{Summary}\

We examined quasineutral plasma equilibria within the framework of a hybrid model treating the electrons as massless fluid satisfying the Boltzmann equation and the ions described by the Vlasov equation. For a cylindrical plasma and spatial dependence on the radial cylindrical coordinate the problem reduces to a set of four quasilenear first-order equations. The respective initial-value problem was solved numerically. The solutions obtained describe either static or stationary steady states with sheared macroscopic ion velocities and associated radial electric fields, thermal electron pressure either constant or peaked on axis and a non gyrotropic ion pressure tensor. Then, the characteristics of a variety of equilibrium solutions were examined for different values of the free parameters involved. The solutions permit the derivation of either diamagnetic or paramagnetic equilibria or equilibria with mixed magnetic properties, i.e. diamagnetic in the inner plasma region and paramagnetic in the outer region. Axial ion-current-density profiles either peaked on-axis or off-axis are possible. Also, the various profiles produced indicate a potential correlation  of the sheared macroscopic ion velocity, the electric field, and the non-gyrotropic element of the ion pressure tensor in consistence with  experimental and theoretical evidence in the improved confinement regimes in the plasmas of fusion devices.

The cylindrical equilibrium solutions of the present study may also describe steady states of magnetic flux-rope structures observed in space plasmas particularly in the magnetosphere \cite{Ng,Yang2014,Xie2023, Vinogradov2016,Eastwood2016,Lui2007,GuGu}. 
%Compared with the steady sates  obtained in \cite{Ng}, those constructed here differ in adopting  the quasineutrality condition and the fact that the cylindrical plasma is radially bounded. 
The study could be extended to one-dimensional equilibria by employing  helical coordinates and to two dimensional equilibria either translational symmetric or axisymmetric or helically symmetric.

%\newpage

\section*{Aknowledgments}\

This work was conducted in the framework of participation of the University of Ioannina in the National Programme for the Controlled
Thermonuclear Fusion, Hellenic Republic.
%The views and opinions expressed herein do not necessarily
%reflect those of the European Commission.

%%%%%%%%%%%%%%%%%%%%%%%%%%%%%%%%%%%%%%%%%%%%%%%%%%%%%%%%%%%%%%%%%%%%%%%%%%%%%%%%%%%%%%%%%%%%%%%%%%%%%%%%%%%%%%%%

\end{document}